8-2019

# Testing Lotka's Law and Pattern of Author Productivity in the Scholarly Publications of Artificial Intelligence

Muneer Ahmad

Sadik Batcha M

# Testing Lotka's Law and Pattern of Author Productivity in the Scholarly Publications of Artificial Intelligence

**Muneer Ahmad[1], Dr. M. Sadik Batcha[2], S Roselin Jahina[3]**

[1]*Ph.D Research Scholar, Department of Library and Information Science, Annamalai University, Tamil Nadu, India , muneerbangroo@gmail.com*
[2]*Research Supervisor and Mentor, Associate Professor, Department of Library and Information Science, Annamalai University, Tamil Nadu, India, msbau@rediffmail.com*
[3]*Ph.D Research Scholar, Department of Library and Information Science, Annamalai University, Tamil Nadu, India , rosejsshaki@gmail.com*

**Abstract**

Artificial intelligence has changed our day to day life in multitude ways. AI technology is rearing itself as a driving force to be reckoned with in the largest industries in the world. AI has already engulfed our educational system, our businesses and our financial establishments. The future is definite that machines with artificial intelligence will soon be captivating over trained manual work that now is mostly cared by humans. Machines can carry out human-like tasks by new inputs as artificial intelligence makes it possible for machines to learn from experience. AI data from web of science database from 2008 to 2017 have been mapped to depict the average growth rate, relative growth rate, contribution made by authors in the view of research productivity, authorship pattern and collaboration of AI literature. The Lotka's law on authorship productivity of AI literature has been tested to confirm the applicability of the law to the present data set. A K-S test was applied to measure the degree of agreement between the distribution of the observed set of data against the inverse general power relationship and the theoretical value of $\alpha = 2$. It is found that the inverse square law of Lotka follow as such.

**Introduction**

The term 'Scientometrics' is a field which consists of the quantitative methods applied to the study of science as an information process, unlike the behavioural sciences and mainstream philosophy of science, it focuses on texts (documents) as empirical units of analysis. It is a scientific discipline, which performs reproducible measurements of scientific activity, and reveals its objective quantitative regularities. Further, Scientometric methods include statistical and thesaurus methods, and indicators as to the number of citations, terms etc. According to Pouris (1989) 'Scientometrics is for science what econometrics is for economics.' Therefore, it is 'Application of quantitative techniques (systems analysis, mathematical and statistical techniques etc.) to scientific communication (science output, science policy, science administration etc.)' with the objectives of developing science indicators; measuring the impact of science on society; and comparing the output as well as the impact of science at national and international levels.

**Artificial Intelligence**

The name behind the idea of AI is John McCarthy, who began research on the subject in 1956 and assumed that each aspect of learning and other domains of intelligence can be described so precisely that they can be simulated by a machine. Artificial intelligence describes the work processes of machines that would require intelligence if performed by humans. The term 'artificial intelligence' thus means 'investigating intelligent problem-solving behaviour and creating intelligent computer systems'. According to the sophistication, AI system can perform action such as perception, interpretation, reasoning, learning, communication and decision making similar to human being to arrive solution for the given problem. The information-processing units in artificial neural networks are artificial neurons similar to the neurons in the human brain (Haykin, 2005). Neural networks learn by experience; that is, they generalize from previous experiences to new ones, and make decisions using those experiences. A neural network consists of a group of neural nodes that are linked to some weighted nodes. Each node simulates a brain neuron, and the connections among these nodes are analogous to the synapses that connect brain neurons.

From the inception, various development have been done on AI system, which broaden its application such as pattern recognition, automation, computer vision, virtual reality, diagnosis, image processing, nonlinear control, robotics, automated reasoning, data mining, process planning, intelligent agent and control, manufacturing, etc. Currently, most applications of AI are narrow, in that they are only able to carry out specific tasks or solve pre-defined problems. AI works in a range of ways, drawing on principles and tools, including from maths, logic, and biology. An important feature of contemporary AI technologies is that they are increasingly able to make sense of varied and unstructured kinds of data, such as natural language text and images. Machine-learning has been the most successful type of AI in recent years, and is the underlying approach of many of the applications currently in use. Rather than following pre-programmed instructions, machine learning allows systems to discover patterns and derive its own rules when it is presented with data and new experiences.

**Review of Literature**

There has been added lot of literature to the field of scientometrics at national (Ahmad, Batcha, Wani, Khan, & Jahina, 2018) (Batcha & Ahmad, 2017) as well as international level (Ahmad & Batcha, 2019; Batcha & Chaturbhuj, 2019). Lotka's Law named after Alfred J. Lotka, describes the frequency of publication by authors in any given field (Lotka, 1926). Bradford law of Scattering describes how the literature on a particular subject is scattered or distributed in the journals (Bradford, 1950). Zipf's law named after the linguist George Kingsley Zipf, states that a given a large sample of words used, the frequency of any word is inversely proportional to its rank in the frequency table (Zipf, 1949). These three laws are the fundamentals of bibliometrics and scientometrics.

Lotka's Law has been applied to the literature of various disciplines by various authors since its publication.

Lotka's Law of authorship productivity is good for application in the field of dentistry literature. The distribution frequency of the authorship follows the exact Lotka's Inverse law with the exponent á=2. The modified form of the inverse square law, i.e., Inverse power law with á and C parameters as 2.49 and 0.733 for dentistry literature is applicable and appears to provide a good fit (Batcha, 2018).

Kumar (2010) examines the applicability of Lotka's Law as a general inverse power ($\alpha \neq 2$) and an inverse square power relationship ($\alpha = 2$) to the distribution of the research productivity in Council of Scientific and Industrial Research (CSIR), India. The results obtained in this study do not follow the inverse square law of Lotka as such and similarly (Gupta, Kumar, & Aggarwal, 1999) has also described in his studies based on CSIR samples that Lotka's formulation is not applicable in case of CSIR productivity distribution. It may be due to longer period of participation in research.

Since the publication of Lotka's paper, numerous authors have attempted to apply Lotka's Law to the literature of various disciplines. While in some studies Lotka's inverse square law holds (e.g., Murphy, 1973) in humanities and (Schorr, 1975) in map librarianship), in others it does not. (Voos, 1974) finds that for the information science literature a new exponent of 3.5 gives the best fit with empirical data. (Schorr, 1974) finds that Lotka's inverse square law is not applicable to the literature of library science and proposes an inverse quadruple law whereby, for each 100 contributors of a single article, about six will contribute two papers, about one will contribute three papers, etc. (Worthen, 1978) reports that Lotka's Law does not fit the literature in medicine. (Radhakrishnan & Kernizan, 1979) find that Lotka's Law does not apply well to computer science literature. They find that an exponent of 3 gives the best fit. In a subsequent study, however, (Subramanyam, 1979) argues that the computer science literature does confirm to Lotka's inverse square law if data are taken from a large collection of journals.

(Batcha, 2018) study lights on Lotka's empirical law of scientific productivity that is inverse square law to measure the scientific productivity of authors, to test Lotka's exponent value and the K.S test for the fitness of Lotka's law and the result obtained in this study do not follow the inverse square law.

(Budd, 1988) applied the Lotka's and Bradford's laws to citations to journals in 569 papers on higher education it finds the conformity of higher education literature, as represented but the database used, is not perfect with the two bibliometric laws, but the results do suggest that the underlying concepts of the laws may have applicability to examination of the discipline.

(Ahmed & Rahman, 2009) examined the validity of Lotka's Law to authorship distribution in the field of nutrition research in Bangladesh (1972-2006) using both generalized and modified models. The results suggest that author productivity distribution predicted in Lotka's generalized inverse square law is not applicable to nutrition research in Bangladesh. While, using LLS method excluding highly productive authors, Lotka's Law was found to be applicable to nutrition research in Bangladesh.

(Aswathy & Gopikuttan, 2013) assessed the author productivity in the publication of three Universities in Kerala during 2005-2009 and (Sudhier, 2013) evaluated the authorship distribution in physics literature. In both the study, Lotka's inverse square law has been applied using Pao's method and the data set was tested by K-S goodness-of-fit-test. But, the Lotka' generalised law is not applicable to these study.

(Naqvi & Fatima, 2017) analysed international business literature to study the applicability of Lotka's law to author productivity. Further, Kolmogorov –Smirnov goodness of fit test (K-S Test) and Chi square test also tested to compare and confirm the dataset. In both the cases, Lotka's law confirmed the author productivity distribution.

**Objectives of the Study**

The objectives of the present study are as follows

- To quantify the research output in the form of publications and average growth rate of literature in the field of Artificial Intelligence literature over the study period of ten years (2008-2017).
- To analyse the authorship pattern and degree of Collaboration of research in the field of Artificial Intelligence literature during the period of study.
- To analyse the research trend with Collaborative Co-efficient, Modulated Collaborative Co-efficient and Collaborative Index in the global literature of Artificial Intelligence.
- To study the growth trend with the analysis of Relative Growth Rate (RGR) of publications;
- To find out the Doubling Time (DT) for the publications to become double of the existing amount;
- To test the applicability of Lotkas's Law in the author productivity.
- To determine whether the "n" value confirms to Lotka's Law through K- S test.

**Methodology**

The data for this study was downloaded from Web of Science database for recent ten years periods, 2008 to 2017 (till October 2018). The data was downloaded on 25-1-2018. The searches were performed on the name of Artificial Intelligence using basic search at the Web of Science Core Collection with all probabilities and bibliographical details of 9734 research papers collectively contributed by 31803 scientists published in 2964 scientific periodicals were collected for Scientometric analysis. Contributions of Indian scientists occupied 6[th] place in order contributing 546 articles.

**Analysis and Interpretation of the Results**

Table 1 depicts the growth of research publications published in the field of Artificial Intelligence literature during the study period 2008-2017.Altogether 9734 publications were published. The highest number of articles, 2085 (21.42%) were published in the year 2017; while less number of research articles 527 (5.41%) were published in the year 2008.The second highest number of articles, 1475 were published in the year 2016 (15.15%). There is a steady growth found in research output from one year to proceeding next year. Further it is found that the average rate of increase in the number of publications per year is 0.862%.

Table 1: Year Wise Distribution and Average Growth Rate of Publications in Artificial Intelligence Literature

| Year | Res.Output | % | Cum.Output | Cum% | Growth Rate |
|------|------------|--------|------------|-------|-------------|
| 2008 | 527 | 5.41% | 527 | 5.41 | |
| 2009 | 600 | 6.66% | 1127 | 11.58 | 0.878 |
| 2010 | 606 | 6.23% | 1733 | 17.80 | 0.990 |
| 2011 | 678 | 6.97% | 2411 | 24.77 | 0.894 |
| 2012 | 763 | 7.84% | 3174 | 32.61 | 0.889 |
| 2013 | 834 | 8.57% | 4008 | 41.18 | 0.915 |
| 2014 | 950 | 9.76% | 4958 | 50.93 | 0.878 |
| 2015 | 1216 | 12.49% | 6174 | 63.43 | 0.781 |
| 2016 | 1475 | 15.15% | 7649 | 78.58 | 0.824 |
| 2017 | 2085 | 21.42% | 9734 | 100 | 0.707 |
| Total | 9734 | 100.00% | | | 0.862 |

Table 2 illustrates the year wise distribution of authorship pattern of global Artificial Intelligence literature. Out of 9734 papers, the authorship pattern up to 10 authors results a total of 9618 research output remaining 116 papers have been published by more than ten authors. Single author contributions are accounted to 17.17% during the study period. The highest percentage of 24.60 is recorded by three authors followed by two and four authors showing 2.97 and 17.39 percentages respectively. However, more than five authors have contributed less than 10 percentages in this study. This analysis of results shows that individual contribution is not at the rate of appreciation compared to collaborative research up to five in the field of artificial Intelligence literature research. The number of authors engaging collaborative research is found increasing year by year from 2008 to 2017 ranging from 1481 to 6973. It can be noticed that 3.31 % f authors/scientists collectively contribute one paper in the field of Artificial Intelligence literature.

Table 2: Analysis of Authorship Pattern among the Scientists of Artificial Intelligence Literature

| Year | 1 | 2 | 3 | 4 | 5 | 6 | 7 | 8 | 9 | 10 | Total | % | Total Authors |
|------|-----|-----|-----|-----|-----|-----|-----|-----|-----|-----|-------|-------|---------------|
| 2008 | 110 | 158 | 124 | 64 | 40 | 18 | 5 | 3 | 2 | 1 | 525 | 5.46 | 1481 |
| 2009 | 113 | 162 | 148 | 105 | 32 | 15 | 13 | 7 | 2 | 2 | 599 | 6.23 | 1748 |
| 2010 | 105 | 156 | 157 | 87 | 44 | 23 | 12 | 9 | 4 | 3 | 600 | 6.24 | 1896 |
| 2011 | 116 | 163 | 178 | 123 | 53 | 19 | 8 | 6 | 2 | 3 | 671 | 6.98 | 2102 |
| 2012 | 113 | 173 | 207 | 150 | 59 | 23 | 14 | 9 | 4 | 3 | 755 | 7.85 | 2460 |
| 2013 | 100 | 180 | 218 | 177 | 81 | 33 | 19 | 12 | 5 | 3 | 828 | 8.61 | 2811 |
| 2014 | 127 | 194 | 227 | 206 | 97 | 44 | 23 | 11 | 8 | 3 | 940 | 9.77 | 3287 |
| 2015 | 199 | 259 | 316 | 208 | 126 | 50 | 24 | 7 | 11 | 1 | 1201 | 12.49 | 4012 |
| 2016 | 225 | 330 | 370 | 238 | 129 | 76 | 37 | 24 | 14 | 11 | 1454 | 15.12 | 5033 |
| 2017 | 443 | 434 | 421 | 315 | 200 | 100 | 63 | 29 | 27 | 13 | 2045 | 21.26 | 6973 |
| Total | 1651 | 2209 | 2366 | 1673 | 861 | 401 | 218 | 117 | 79 | 43 | 9618 | 100 | 31803 |
| % | 17.17 | 22.97 | 24.60 | 17.39 | 8.95 | 4.17 | 2.27 | 1.22 | 0.82 | 0.45 | 100.00 | AAPP* | 3.31 |

*AAPP – Average author per paper

**Collaborative Index (CI)**

(Lawani,1986) proposed and coined the term Collaborative Index to describe the average number of authors per paper for a given set of papers and used it as a quantitative measure of research collaboration. It can be calculated easily, but it cannot be interpreted as a degree because it has no upper- value limit. It is denoted by the formula:

$$C = \frac{\text{Total Number of Authors}}{\text{Total Number of Papers}}$$

**Degree of Collaboration (DC)**

Subramanyam propounded the DC, a measure to calculate the proportion of single and multi-author papers and to interpret it as a degree. According to Subramanyam (1983),

$$DC = \frac{Nm}{Ns+Nm} \qquad \frac{\text{No of Muti-authored papers}}{\text{No of Single+No of Multi-authored Papers}}$$

**Co-Authorship Index (CAI)**

To study the shift in the pattern of co-authorship during 2008-2017 CAI suggested by (Garg & Padhi, 2001) was used

CAI is computed as follows

$$CAI = \left[ \frac{\left(\frac{N_{ij}}{N_{io}}\right)}{\left(\frac{N_{oj}}{N_{oo}}\right)} \right] \times 100$$

Where Nij: number of papers having j authors in year/block;

Nio : total output of year block i;

Noj : number of papers having j authors for all years/blocks;

Noo : total number of papers for all authors and all years/blocks.

J  =  2, (3 or 4), >= 5

**Collaboration Co-efficient (CC)**

(Ajiferuke et al., 1988) suggested a single measure to measure collaborative research and termed it as collaborative co-efficient. The method is based on fractional productivity defined by Price and Beaver .The following formula denotes CC. The symbols used have been explained as under:

$$CC = 1 - \frac{\sum_{j}^{k}(1/j)f_{j}}{N}$$

Where $f_j$ is the number of j authored papers; N is the total number of research papers published and k is the greatest number of authors per paper according to Ajiferuke, CC tends to zero as single authored papers dominate and to 1-1/j as j-authored papers dominate. This implies that higher the value of CC, higher the probability of papers with multi or mega authors.

**Modified Collaboration Co-efficient (MCC)**

(Sayanur and Srikanth, 2010) modified the CC and derived MCC as follows:

$$MCC = \frac{A}{A-1}1 - \frac{\sum_{j}^{k}(1/j)f_{j}}{N}$$

Table 3 attempts to analyse different collaboration factors for the period of 10 years (2008-2017). The analysis of the table includes CI, DC, CAI, CC, and MCC. The Table shows collaborative Index at the lowest level in the year 2008. Collaborative Index is highest in the year 2014 and mean CI during the time of study is 3.24. Subramanyam propounded the DC, a measure to calculate the proportion of single and multi-author papers and to interpret it as a degree. DC varies from 0 when all the papers have a single author to 1 when all the papers have more than one author. It can be easily calculated and can also be easily interpreted.

Table 3: Analysis of Collaboration factors in Artificial Intelligence Publication at Global Level

| Authorship Pattern | 2008 | 2009 | 2010 | 2011 | 2012 | 2013 | 2014 | 2015 | 2016 | 2017 | Total |
|---|---|---|---|---|---|---|---|---|---|---|---|
| 1 | 110 | 113 | 105 | 116 | 113 | 100 | 127 | 199 | 225 | 443 | 1651 |
| 2 | 158 | 162 | 156 | 163 | 173 | 180 | 194 | 259 | 330 | 434 | 2209 |
| 3 | 124 | 148 | 157 | 178 | 207 | 218 | 227 | 316 | 370 | 421 | 2366 |
| 4 | 64 | 105 | 87 | 123 | 150 | 177 | 206 | 208 | 238 | 315 | 1673 |
| 5 | 40 | 32 | 44 | 53 | 59 | 81 | 97 | 126 | 129 | 200 | 861 |
| 6 | 18 | 15 | 23 | 19 | 23 | 33 | 44 | 50 | 76 | 100 | 401 |

| 7 | 5 | 13 | 12 | 8 | 14 | 19 | 23 | 24 | 37 | 63 | 218 |
| 8 | 3 | 7 | 9 | 6 | 9 | 12 | 11 | 7 | 24 | 29 | 117 |
| 9 | 2 | 2 | 4 | 2 | 4 | 5 | 8 | 11 | 14 | 27 | 79 |
| 10 | 1 | 2 | 3 | 3 | 3 | 3 | 3 | 1 | 11 | 13 | 43 |
| Total Paper | 525 | 599 | 600 | 671 | 755 | 828 | 940 | 1201 | 1454 | 2045 | 9618 |
| Total Author | 1481 | 1748 | 1896 | 2102 | 2460 | 2811 | 3287 | 4012 | 5033 | 6973 | 31803 |
| CI* | 2.82 | 2.92 | 3.16 | 3.13 | 3.26 | 3.39 | 3.50 | 3.34 | 3.46 | 3.41 | 3.24 |
| DC+ | 0.79 | 0.81 | 0.83 | 0.83 | 0.85 | 0.88 | 0.86 | 0.83 | 0.85 | 0.78 | 0.83 |
| CAI! | 95.43 | 97.95 | 99.60 | 99.85 | 102.65 | 106.14 | 104.41 | 100.72 | 102.04 | 94.57 | 100.34 |
| CC# | 0.3766 | 0.4206 | 0.4479 | 0.4526 | 0.493 | 0.5416 | 0.528 | 0.476 | 0.4937 | 0.4053 | 0.4635 |
| MCC$ | 0.3775 | 0.4215 | 0.4488 | 0.4534 | 0.4938 | 0.5423 | 0.5287 | 0.4765 | 0.4941 | 0.4056 | 0.4642 |
| MCC-CC | 0.0009 | 0.0009 | 0.0009 | 0.0008 | 0.0008 | 0.0007 | 0.0007 | 0.0005 | 0.0004 | 0.0003 | 0.0007 |

CI* - Collaborative Index, DC+ - Degree of Collaboration, CAI! – Co-authorship Index, CC# - Collaborative Co-efficient, MCC$ - Modified Collaborative Co-efficient

It is found in this study that DC was lowest at 0.78 in 2017 and highest at 0.88 in 2013. In all the year multi-authored papers are steadily increasing, but in 2017 it is at its lowest and hence the mean DC during the study period shows 0.83. The value of CAI in the first year starts with 95.43 and it progressively increases in respect of other proceeding years as multi and mega authored papers increase. This implied that during the first year of study single authored papers are high at the scenario. The year 2008 onwards the value of CAI steadily increases from 95.43 to 106.14 in 2013 and from 2014 to goes down further and in year 2017 is goes lowest i.e. 94.57 suggesting the trend in the later years is marked with less research papers with small team size. This result is supported with the outcome of CC. In this study, CC is also lowest in 2008 showing 0.3766. It is at the highest rate of 0.5416 in 2013. The mean CC is 0.4635.

The study found MCC was lowest in 2008, when it was 0.3775. It was at the highest value of 0.5423 in 2013. The mean MCC during the period of study was 0.4642. It is also observed from the table that the mean difference between CC and MCC is 0.0007. Least difference between CC and MCC, i.e. 0.0003 is observed during the year 2017. The highest difference CC and MCC, which is 0.0009, is observed in the years 2008, 2009 and 2010. It can be concluded that no significant difference can be observed between CC values and MCC values, and also this variation narrows down when the number of authorships increases.

Out of 9618 articles published, single author share is 1651 and multiple paper author shares is 7967. This indicates that single paper contribution is less than multiple author papers. It can be summarized from the above discussion that very high collaborative research activities are observed in global artificial intelligence literature.

**Lotka's Law**

Lotka's Law is one of the most basic laws of bibliometrics, which deals with the frequency of publication of authors in and given field. The generalized form of Lotka's law can be expressed as

Y = (C)

Where Y is the number of authors with X articles, the exponent n and constant C are parameters to be estimated from a given set of author productivity data.

Lotka's law describes the frequency of publication by authors in a given field. It states that the number of authors making n contribution is about $1/n^2$ on those making one and the proportion of all contributions that make a single contributions, is about 60 percent (Potter, 1981). This means that out of all the authors in a given field, 60 percent will have just one publication and 15 percent will have two publications. 7 percent of authors will have three publications and so on. According to Lotka's law of scientific productivity, only six percent the authors in a field will produce more than 10 articles.

While theoretical Lotka's value is á = 2.00

Theoretical value of 'n' 2.84 is matched with table value of R. Rosseau for getting C.S. value 0.8083

| Constant Value of Present Study | n Value |
|---|---|
| 0.8083 | 2.84 |
| Lotka's Constant Value | n Value |
| 0.6079 | 2 |

D-Max Value Present Study        D-Max Value of Lotka's Study

    0.0073                                        0.1314

To test the goodness of fit, weather the observed author productivity distribution is not significantly different from theoretical distribution. K-S test applied. According to this test, the maximum deviation is observed and estimated value D-Max is calculated follows:

$D_{Max} = F(x) - En(x)$  á = 2.84

Theoretical Value of $C = 0.8083$ Fe+ $= 0.8083 \left( \dfrac{1}{x^{2.84}} \right)$

D-Max = 0.0073

Critical Value at 0.01 level of significance $= \dfrac{2.84}{\sqrt{23460}} = 0.0185$

Table 4: Analysis of Lotkas's Exponent Value on Artificial Intelligence Research Output

| x | y | X=Log X | Y=Log Y | XY | X² |
|---|---|---|---|---|---|
| 1 | 18995 | 0.000000000 | 4.278639298 | 0.00000000 | 0.000000000 |
| 2 | 2826 | 0.301029996 | 3.451172158 | 1.03890634 | 0.090619058 |
| 3 | 860 | 0.477121255 | 2.934498451 | 1.400111583 | 0.227644692 |
| 4 | 344 | 0.602059991 | 2.536558443 | 1.527160354 | 0.362476233 |
| 5 | 167 | 0.698970004 | 2.222716471 | 1.553612141 | 0.488559067 |
| 6 | 83 | 0.778151250 | 1.919078092 | 1.493333017 | 0.605519368 |
| 7 | 46 | 0.845098040 | 1.662757832 | 1.405193385 | 0.714190697 |
| 8 | 48 | 0.903089987 | 1.681241237 | 1.518312127 | 0.815571525 |
| 9 | 26 | 0.954242509 | 1.414973348 | 1.350227718 | 0.910578767 |
| 10 | 65 | 1.000000000 | 1.812913357 | 1.812913357 | 1.000000000 |
|  |  | 6.559763033 | 23.91454869 | 13.09977002 | 5.215159407 |
|  |  | ΣX | ΣY | ΣXY | ΣX² |

The theoretical value of C as 0.8083 for á = 2.84 is taken from the book 'Power Laws in the Information Production Process: Lotkaian Informetrics' by Egghe (2005). The K-S test is applied for the fitness of Lotka's law fits to the global Artificial Intelligence research output. Result indicates that the value of D- Max, i.e. 0.0073 determined with Lotka's exponent á = 2.84 for artificial intelligence which is not close and shows high to the D-Max value 0.2018 determined with the Lotka's exponent á = 2 than the critical value determined at the 0.01 level of significance, i.e., 0.0185. Thus, distribution frequency of the authorship follows the exact Lotka's inverse law with the exponent á = 2. The modified form of the inverse square law, i.e., inverse power law with á and C parameters as 2.84 and 0.8083 for artificial intelligence literature is applicable and appears to provide a good fit.

Table 5: K-S Test of Observed and Expected Distribution of Authors

| x | y | Observed= yx/Σxy | Value=Σ (yx/Σyx) | Expected Freq | Value of Freq/Cum | Diff(D) | Expected Freq | Value of Freq/Cum | Diff(D) |
|---|---|---|---|---|---|---|---|---|---|
| 1 | 18995 | 0.8097 | 0.8097 | 0.8083 | 0.8083 | 0.0014 | 0.6079 | 0.6079 | 0.2018 |

| 2 | 2826 | 0.1205 | 0.9301 | 0.1131 | 0.9214 | 0.0073 | 0.1520 | 0.7599 | -0.0315 |
| --- | --- | --- | --- | --- | --- | --- | --- | --- | --- |
| 3 | 860 | 0.0367 | 0.1571 | 0.0358 | 0.9572 | 0.0008 | 0.0675 | 0.8274 | -0.0308 |
| 4 | 344 | 0.0147 | 0.0513 | 0.0158 | 0.9731 | -0.0012 | 0.0380 | 0.8654 | -0.0233 |
| 5 | 167 | 0.0071 | 0.0218 | 0.0084 | 0.9815 | -0.0013 | 0.0243 | 0.8897 | -0.0172 |
| 6 | 83 | 0.0035 | 0.0107 | 0.0050 | 0.9865 | -0.0015 | 0.0169 | 0.9066 | -0.0134 |
| 7 | 46 | 0.0020 | 0.0055 | 0.0032 | 0.9897 | -0.0013 | 0.0124 | 0.9190 | -0.0104 |
| 8 | 48 | 0.0020 | 0.0040 | 0.0022 | 0.9919 | -0.0002 | 0.0095 | 0.9285 | -0.0075 |
| 9 | 26 | 0.0011 | 0.0032 | 0.0016 | 0.9935 | -0.0005 | 0.0075 | 0.9360 | -0.0064 |
| 10 | 65 | 0.0028 | 0.0039 | 0.0012 | 0.9947 | 0.0016 | 0.0061 | 0.9421 | -0.0033 |
| Total | 23460 | | Present study's | | D.Max = 0.0073 | | Lotka's | | D.Max = 0.2018 |

**Relative Growth Rate (RGR)**

Relative Growth Rate (RGR) means the increase in the number of articles per unit of time. The mean RGR of articles over the specific period of interval which is mathematically given by:

Rt(P) = [logP(t)-logP(0)]

Rt = Relative growth rate of articles over the specific period of time.

LogP (0) = Logarithm of initial number of articles logP (t)

= Logarithm of final number of articles.

**Doubling Time**

Doubling time is defined as the time required for the articles to become double of the existing amount. It has been calculated using following formula:

Dt is given by (t) =0.693/R

Where R is relative growth rate of articles

Dt = It is directly related to RGR.

Table 6 clearly indicates the average Relative Growth Rate Rt(P) and Doubling time of articles in artificial intelligence literature during the study period. It is observed that the value of relative growth rate of an article has gradually increased from 2008 (-0.0002) to 2017 (1.5405) and in that order the value of doubling time of the articles Dt(P) gradually decreased from 1.0998 year (2008) to 0.4499 year (2017). The mean relative growth rate Rt(P) of articles for the first five years (from 2008 to 2012) was 0.52262. It increased to 1.60662 for the next five years (from 2013 to 2017), whereas for the doubling time of the articles Dt(P) for the first five years (from 2008 to 2012) indicates 0.912275 has gradually decreased for the next five years (from 2013 to 2017) to 0.43168. It can be concluded from the above analysis that relative growth rate of articles has been gradually increased and on the other hand, doubling time of the articles has been gradually decreased.

Table 6: Relative Growth Rate and Doubling Time of Global Artificial Intelligence Literature

| Year | Res.Output | Cum.Output | W1 | W2 | RT(p) | Mean RP(p) | Dt(p) | Mean Dt(p) |
|------|------------|------------|------|------|---------|------------|--------|------------|
| 2008 | 527 | 527 | 6.2672 | 6.267 | -0.0002 | | | |
| 2009 | 600 | 1127 | 6.3969 | 7.027 | 0.6301 | | 1.0998 | |
| 2010 | 606 | 1733 | 6.4069 | 7.458 | 1.0511 | 0.52262 | 0.6593 | 0.912275 |
| 2011 | 678 | 2411 | 8.2816 | 7.788 | -0.4936 | | 1.4039 | |
| 2012 | 763 | 3174 | 6.6373 | 8.063 | 1.4257 | | 0.4861 | |
| 2013 | 834 | 4008 | 6.7262 | 8.296 | 1.5698 | | 0.4415 | |
| 2014 | 950 | 4958 | 6.8565 | 8.509 | 1.6525 | | 0.4194 | |
| 2015 | 1216 | 6174 | 7.1033 | 8.728 | 1.6247 | 1.60662 | 0.4265 | 0.43168 |
| 2016 | 1475 | 7649 | 7.2964 | 8.942 | 1.6456 | | 0.4211 | |
| 2017 | 2085 | 9734 | 7.6425 | 9.183 | 1.5405 | | 0.4499 | |
| Total | 9734 | | | | | | | 0.6719775 |

**Conclusion**

The study quantitatively identified the research productivity in the area of artificial intelligence at global level over the study period of ten years (2008-2017). The study identified the trends and characteristics of growth and collaboration pattern of artificial intelligence research output. Average growth rate of artificial intelligence per year increases at the rate of 0.862. The multi-authorship pattern in the study is found high and the average number of authors per paper is 3.31. Collaborative Index is noted to be the highest range in the year 2014 with 3.50. Mean CI during the period of study is 3.24. This is also supported by the mean degree of collaboration at the percentage of 0.83 .The mean CC observed is 0.4635. Lotka's Law of authorship productivity is good for application in the field of artificial intelligence literature. The distribution frequency of the authorship follows the exact Lotka's Inverse Law with the exponent á = 2. The modified form of the inverse square law, i.e., Inverse Power Law with á and C parameters as 2.84 and 0.8083 for artificial intelligence literature is applicable and appears to provide a good fit. Relative Growth Rate [Rt(P)] of an article gradually increases from -0.0002 to 1.5405, correspondingly the value of doubling time of the articles Dt(P) decreases from 1.0998 to 0.4499 (2008-2017). At the outset the study reveals the fact that the artificial intelligence literature research study is one of the emerging and blooming fields in the domain of information sciences.